\documentclass[aps,pre,superscriptaddress,twocolumn,amsmath,amssymb]{revtex4}
\usepackage{amsmath,amsfonts,amssymb,graphicx}
\usepackage{epsfig}
\usepackage{epstopdf}
\newcommand{\be}{\begin{equation}}
\newcommand{\ee}{\end{equation}}
\newcommand{\bey}{\begin{eqnarray}}
\newcommand{\eey}{\end{eqnarray}}
\newcommand{\bw}{\begin{widetext}}
\newcommand{\ew}{\end{widetext}}

\newcommand{\ov}{\overline}

\newcommand{\ra}{\rangle}
\newcommand{\la}{\langle}

\newcommand{\ba}{\begin{array}}
\newcommand{\ea}{\end{array}}
\newcommand{\bi}{\begin{itemize}}
\newcommand{\ei}{\end{itemize}}
\newcommand{\bem}{\begin{enumerate}}
\newcommand{\eem}{\end{enumerate}}

\newcommand{\hefei}{Department of Modern Physics, University of Science and Technology of China, Hefei 230026, China}

\newcommand{\xiamen}{Department of Physics, Fujian Key University Laboratory for Low-Dimensional Condensed Matter Physics, and Jiujiang Research Institute, Xiamen University, Xiamen 361005, Fujian, China}

\begin{document}
\title{Thermalization of small quantum systems: From the zeroth law of thermodynamics}

\author{Jiaozi Wang}
\affiliation{\hefei}
\author{Wen-ge Wang}
\email{wgwang@ustc.edu.cn}
\affiliation{\hefei}
%\author{Giuliano Benenti}
%\affiliation{\como}
%\affiliation{\infn}
%\affiliation{\NEST}
%\author{Giulio Casati}
%\affiliation{\como}
%\affiliation{\brazil}
\author{Jiao Wang}
\affiliation{\xiamen}
\date{\today}

\begin{abstract}
Thermalization of isolated quantum systems has been studied intensively in recent years and significant progresses have been achieved. Here, we study  thermalization of small quantum systems that interact with large chaotic environments under the consideration of Schr\"{o}dinger evolution of composite systems, from the perspective of the zeroth law of thermodynamics. Namely, we consider a small quantum system that is brought into contact with a large environmental system; after they have relaxed, they are separated and their temperatures are studied. Our question is under what conditions the small system may have a detectable temperature that is identical with the environmental temperature. This should be a necessary condition for the small quantum system to be thermalized and to have a well-defined temperature. By using a two-level probe quantum system that plays the role of a thermometer, we find that the zeroth law is applicable to quantum chaotic systems, but not to integrable systems.
\end{abstract}

\maketitle

\section{Introduction}
Whether and how a quantum system reaches thermalization is a long-standing problem that has been perseveringly
studied for long~\cite{Qtherm_Haar55, Qtherm_Peres84, Qtherm_Shankar79, Qtherm_Tasaki98, Qtherm_typ_Popescu06, Qtherm_typ_Goldstein06}, since von Neumann's seminal work~\cite{Qtherm_Neuann}. Modern experimental advances in ultracold atoms and trapped ions~\cite{Exp_Kinoshita06, Exp_Gring12, Exp_Gross12, Exp_Bloch12, Exp_Eisert12, Exp_Langen13, Exp_Greiner16, Exp_Richerme14, Exp_Clos16} have attracted more interest and efforts on this problem~\cite{review_Polkovnikov16, review_Eisert15, review_Polkovnikov11, review_Rigol16}, with particular focus on small quantum systems~\cite{review_Polkovnikov16}. Generally speaking, studies on this issue proceeded in two directions: one for isolated systems and another for systems coupled with environments.

For isolated quantum systems, rapid and important progresses have been achieved in recent years~\cite{review_Izrailev16, review_Rigol16}. The eigenstate thermalization hypothesis (ETH), first proposed in Refs.~\cite{ETH_Deutsch}
and~\cite{ETH_Srednicki}, has established a general framework for understanding thermalization results in isolated systems. It claims that for local observables, every energy eigenstate is thermal, i.e., indistinguishable from the corresponding (micro)canonical ensemble~\cite{ETH_Rigol08}. It implies the possibility that the expectation value of a local observable in an isolated quantum system far from equilibrium may relax to a (almost) time independent result that agrees with the ensemble average predicted by orthodox statistical mechanics~\cite{Qtherm_Iso_Rigol12, Qtherm_Iso_Izrailev12, Qtherm_Iso_Sorg14, Qtherm_Iso_Santos14,  Qtherm_Iso_Srednicki99, Qtherm_Iso_Rigol14, ETH_Rigol08}.

But for systems interacting with environments, not as many works have been done and some important problems remain open~\cite{Qtherm_SE_Yuan09, Qtherm_SE_Freeman14, Qtherm_SE_Cramer15, Qtherm_SE_Linden09, Qtherm_SE_Genway12, Qtherm_SE_Eisert12, Qtherm_SE_WWG12, Qtherm_SE_Fiako15,Sed13}. For example, one basic question is under what condition a system may evolve into a steady state and, when this happens, whether the system may be characterized by a temperature. Indeed, although temperature is a core concept for quantum thermalization, it has not been fully understood yet~\cite{Temp_Hanggi14, Temp_local_Eisert14, Temp_NET_Braun13}. In this respect, it was found that, when the effective dimension of the initial state of the system-environment composite is large, the reduced density matrix (RDM) of the system will relax to a steady state~\cite{Qtherm_SE_Linden09}. Under some assumptions~\cite{Qtherm_SE_Genway12, Qtherm_SE_Eisert12, Qtherm_SE_WWG12}, the steady RDM may further result in a Gibbs form parameterized by the environmental temperature. Following these progresses, one immediate task is to justify those assumptions, or to reveal the conditions under which they are applicable. Another crucial task is to clarify and distinguish two temperatures involved here. One we refer to as the \emph{environment-assigned temperature}, denoted by $\beta_E$, which is the temperature of the environment the system interacts with. It also appears as a parameter in the steady RDM of the system after relaxation. Another temperature is the temperature the system may own when it is decoupled from the environment after relaxation. We refer to it as the \emph{internal temperature}, denoted by $\beta_I$; it characterizes the motion of the system in its own internal freedom at the thermal state it may reach.

 One question is, under what condition an internal temperature can be defined and related to the environment-assigned temperature. So far,  internal temperature has not been addressed in the context of environment interaction and compared with the environment temperature, yet. In literatures, explicitly or implicitly, the two temperatures have been assumed to be identical, making the meaning of temperature somehow confusing. One reason could be that $\beta_E$ appears in the Gibbs form of the RDM.
 But, the role played by the RDM is only the `initial condition' when being disconnected from the environment, which cannot fully govern the system's ongoing evolution and hence its thermal property.
 Another reason might be that the zeroth law of thermodynamics has been taken for granted in the present context. Nevertheless, it is well known that in principle the thermodynamic laws are for macroscopic systems exclusively. The above question is equivalent to ask under what condition or to what extent the zeroth law can be extended to small quantum systems. Our motivation in this work is to clarify this question.

To this end, a necessary and important step is to measure and determine the temperature of a studied small quantum system when it is decoupled from the environment, if it may thermalize. As the system is isolated when being decoupled, we can take advantage of the temperature-detecting strategy developed recently in Ref.~\cite{WW-PRE17} for an isolated quantum system based on the  principle that, as an internal property of the system itself, internal temperature can only be assigned to the system when it can be detected in a reliable way and the measured value is independent of detection details.

The rest of the paper is organized as follows. In Sec.~\ref{sect-ECT}, we will study the
Schr\"{o}dinger evolution of the system-environment composite and show that, regardless of the size and the dynamics of the system, the RDM will relax to a Gibbs steady state given by $\beta_E$ in general. In Sec.~\ref{sect-INT}, we will study the measurement of $\beta_I$ and reveal the condition for that such a measurement can be performed. In Sec.~\ref{sect-numerical}, we will put our theoretical results into numerical tests, and finally, we will summarize our study in Sec.~\ref{sect-concl}.

\section{Relaxation of the system-environment composite}\label{sect-ECT}
In this section, we study thermalization of a small quantum system in contact with a large quantum environment,  with the consideration of Schr\"{o}dinger evolution of composite systems. We will show that, under the conditions stated below on both the environment and the coupling term, a system, chaotic or integrable, will finally evolve into a Gibbs steady state with the corresponding $\beta$ parameter equal to the environment temperature, or $\beta_E$.  The conditions are:
\bi
\item For the environment,
(i) it should be a chaotic system,
(ii) it should be large and its density of states can be approximately assumed to be an exponential function within the energy range of the small quantum system, and (iii) the number of eigenstates in initial state should be also large.
\item For the coupling term, it should be appropriately adjusted to satisfy that
(i) the system-environment composite is chaotic, and
(ii) the width of the local density of states (LDOS) of the composite system is narrow [to satisfy Eq.~(\ref{eq-LDOS-SE}) in Appendix B], which implies that the coupling is relatively weak.
\ei

Let us begin with our main setup. The Hamiltonian of the composite system is
\be
H=H_{S}+\lambda H_{IS}\otimes H_{I{\cal E}}+H_{{\cal E}}
\ee
%its eigenstates are denoted by $|\phi^{\cal E}_j \rangle$ with energy $E^{\cal E}_j$, $H_{{\cal E}}|\phi_{j}^{{\cal E}}\rangle=E_{j}^{{\cal E}}|\phi_{j}^{{\cal E}}\rangle$. $|E^S _k \rangle$ is used to denote eigenstates of $H_S$ with eigenvalues $E^S _k$, $H_{{\cal S}}|\psi_{k}^{{\cal S}}\rangle=E_{k}^{{\cal S}}|\psi_{k}^{S}\rangle$, and we denote the width of spectrum of $H_S$ by $\Delta_{E_S}$.
with $H_S$ and $H_{\cal E}$ being that of the system and of the environment, respectively, and the middle term on the
r.h.s. being the coupling. The eigenstates of $H_S$ and $H_{\cal E}$ are denoted by $|\psi_{k}^{S}\rangle$ and $|\phi^{\cal E}_j \rangle$,
\be
\ H_{{\cal S}}|\psi_{k}^{{\cal S}}\rangle=E_{k}^{{\cal S}}|\psi_{k}^{S}\rangle, \quad H_{{\cal E}}|\phi_{j}^{{\cal E}}\rangle=E_{j}^{{\cal E}}|\phi_{j}^{{\cal E}}\rangle\ ,
\ee
and the corresponding density of states are denoted by $\rho_S(E^S)$ and $\rho_{\cal E}(E^{\cal E})$, respectively. Eigenstates of the total Hamiltonian $H$ are denoted by $|\psi_\alpha\ra$ with energies $E_{\alpha}$, $H|\psi_{\alpha}\rangle=E_{\alpha}|\psi_{\alpha}\rangle$, which can be decomposed as
\begin{equation}
 |\psi_{\alpha}\ra  =  \sum_{k,j}C_{kj}^{\alpha}|\psi^S _{k}\rangle | \phi^{\cal E} _j\rangle.
\end{equation}
The LDOS of an unperturbed state $|\psi^S _{k}\rangle | \phi^{\cal E} _j\rangle$ is defined as $L_{kj}(E) = \sum|C_{kj}^{\alpha}|^{2} \delta(E_{\alpha}-E)$.

%For convenience of further discussion, we define LDOS of unperturbative state $|\psi^S _{k}\rangle | \phi^{\cal E} _j\rangle$ as $\rho_{kj}^{L}(E)=\sum|C_{kj}^{\alpha}|^{2}\delta(E_{\alpha}-E)$.
%For convenience of further discussion, we introduce $\Delta_S$ to denote the width of spectrum of $H_S$,
%$\rho^L _{kj} (E)$ is to denote the LDOS of the composite system with the defination,

According to recent progresses achieved in the study on foundation of quantum statistical mechanics, many properties of the so-called typical states within appropriately-large energy shells are similar to those of thermal states. So, a typical state of the environment, representing a thermal state with a temperature $\beta_0$, is taken as its initial state, which reads
\be
 |\Phi^{\cal E}_0\rangle = \sum_{E^{\cal E}_{j} \in \Gamma_0} D_{j} |\phi^{\cal E} _j\rangle,
\ee
where $D_j$ are Gaussian random numbers with a same variance, with the normalisation condition $\sum_j |D_j|^2=1$. We use $N_{\Gamma_0}$ to indicate the number of energy levels in the energy shell $\Gamma_0 = [E^{\cal E}_{0}-\delta E/2, E^{\cal E}_{0}+\delta E/2]$, and the value of $E^{\cal E}_0$ is determined by $\beta_{0}$ as $\frac{\partial\ln\rho_{{\cal E}}(E)}{\partial E}|_{E=E_{0}^{{\cal E}}}=\beta_{0}$. For the system, we consider a generic initial state $|\psi^S _0 \rangle =\sum_k C_k |\psi ^S _k\rangle$.

Now we study the time evolution of the composite system after the interaction is turned on, with the consideration of Schr\"{o}dinger evolution of composite system.
Note that, due to the previously stated requirement that the environment is large and has an exponential density of states, the interaction does not change the temperature of the environment.

If the composite system is a quantum chaotic system, its energy levels, as well as their spacings, should have no degeneracy. It is known that, in this situation, beyond a sufficiently long time, the distance between the RDM of the system, $\rho_S(t) = {\rm Tr}_{\cal E} (|\Psi(t) \ra \la \Psi(t)|)$, and its long-time average, denoted by $\overline{\rho}_S$, scaling as $N_{\Gamma_0}^{- 1/2}$, where $\overline{\rho}_S$ is defined as
$\overline{\rho}_{S}=\lim_{T\rightarrow\infty}\frac{1}{T}\int_{0}^{T}\rho_{S}(t)dt$.
 This implies that for large $N_{\Gamma_0}$, if $\rho_S(t)$ has a steady state, it should be $\overline{\rho}_S$. Applying results given in Refs.~\cite{pre14-ps,YWW19} to the system-environment composite studied here with ${\rm Tr}_S(H_I)=0$, we can see that the steady state of the system have an approximately-diagonal form in the eigenbasis $\{ |\psi^{S}_k\ra\}$, when the coupling parameter $\lambda$ satisfies the conditions mentioned previously.
In other words, the eigenbasis of the self-Hamiltonian of the system is a preferred basis~\cite{Zurek-ps,pra08-ps}, so we only need to study the diagonal elements $\rho_{kk}(t)=\langle\psi_{k}^{S}|\rho_{S}(t)|\psi_{k}^{S}\rangle$, especially its steady value $\overline{\rho}_{kk}=\langle \psi^S _k| \overline{\rho}_S |\psi^S _k \rangle$. Here we omit the subscript $S$ in $\overline{\rho}_{kk}$ and $\rho_{kk}(t)$ for brevity. Considering the random feature of components of eigenfunctions (EF) in chaotic systems, one can write $\overline{\rho}_{kk}$ as (see Appendix A)
\be\label{eq-rhokk-origin}
\overline{\rho}_{kk}\simeq\frac{1}{N_{\Gamma_0}}\sum_{k_0}|C_{k_0}|^2 \sum_{E_{j_{0}}^{{\cal E}} \in \Gamma_0} |C_{k_{0} j_{0}}^{\alpha}|^{2} \sum_{\alpha}\sum_{j}|C_{kj}^{\alpha}|^2.
\ee
By defining ${\cal F}_{k_{0}j_{0}}^{k} = \sum_{\alpha}|C_{k_{0}j_{0}}^{\alpha}|^{2} F_{\alpha k}$ with $F_{\alpha k}=\sum_{j}|C_{kj}^{\alpha}|^{2}$, $\overline{\rho}_{kk}$ can be rewritten as
\be
\overline{\rho}_{kk}=\frac{1}{N_{\Gamma_{0}}}\sum_{k_{0}}|C_{k_{0}}|^{2}\sum_{E_{j_{0}}^{{\cal E}}\in\Gamma_{0}}{\cal F}_{k_{0}j_{0}}^{k}.
\ee
If the environment is chaotic and the coupling term is adjusted to satisfy the conditions given in the beginning of this section, we can derive the value of $F_{\alpha k}$ and ${\cal F}_{k_{0}j_{0}}^{k}$ (see Appendix B) as
\be\label{eq-Fak}
F_{\alpha k}\simeq F_k (E_\alpha)\simeq\frac{\rho_{{\cal E}}(E_{\alpha}-E_{k}^{S})}{\sum_{l}\rho_{{\cal E}}(E_{\alpha}-E_{l}^{S})},
\ee
and
\be\label{eq-cFkj}
{\cal F}_{k_{0}j_{0}}^{k}\simeq F_{k}(E_{k_{0}}^{S}+E_{j_{0}}^{{\cal E}})\simeq\frac{\rho_{{\cal E}}(E_{k_{0}}^{S}+E_{j_{0}}^{{\cal E}}-E_{k}^{S})}{\sum_{l}\rho_{{\cal E}}(E_{k_{0}}^{S}+E_{j_{0}}^{{\cal E}}-E_{l}^{S})}.
\ee
If $\rho_{\cal E}(E)$ can be taken as an nearly exponential function within the energy range of the small system, then
\be\label{eq-exprho}
\rho_{{\cal E}}(E_{k_{0}}^{S}+E_{j_{0}}^{{\cal E}}-E_{l}^{S})\simeq\rho_{{\cal E}}(E_{k_{0}}^{S}+E_{j_{0}}^{{\cal E}}) \exp(-\beta_{0}E_{l}^{S}).
\ee
Substituting Eq.~(\ref{eq-exprho}) into Eq.~(\ref{eq-cFkj}), we therefore have
\be\label{eq-Fkk}
{\cal F}_{k_{0}j_{0}}^{k}\simeq\frac{\exp(-\beta_0 E_{k}^{s})}{\sum_{k}\exp(-\beta_0 E_{k}^{S})}.
\ee
Noting that  ${\cal F}_{k_{0}j_{0}}^{k}$ is independent of $k_0$ and $j_0$ according to Eq.(\ref{eq-Fkk}) and the normalisation condition $\sum_{k_0}|C_{k_0}|^2=1$, we can further obtain that
\be\label{eq-rhokk}
\overline{\rho}_{kk}\simeq\frac{\exp(-\beta_0 E_{k}^{s})}{\sum_{k}\exp(-\beta_0 E_{k}^{S})}.
\ee

It is worth emphasizing that in the calculation of $\overline{\rho}_{kk}$, we have only assumed that the environment and the composite system are chaotic, without any requirements on the system itself.
This implies that a small quantum system, in spite of its size and whether chaotic or not, should evolve into the Gibbs state when being coupled to a large environment with appropriate coupling.
Though as a parameter the temperature of the environment enters the RDM, we cannot assume for granted that it serves as the internal temperature of the system. We are to study this topic in the next section.

\section{Detection of the temperature of system}\label{sect-INT}
In this section, we study how to determine the temperature of the system after it is decoupled from the environment with a steady RDM, based on the method established in Ref.~\cite{WW-PRE17}. We will show that the system can possess a well-defined temperature only when it is chaotic and its size is not too small, and importantly, in this case the value of the temperature is close to the environment temperature.

First of all, let us recall some main results of Ref.~\cite{WW-PRE17} and explain the concept of internal temperature. Different from the environment temperature, internal temperature $\beta_I$ should be an internal property of system and should be detected in a reliable way.
If we want to detect a small system's temperature by coupling a probe (thermometer) to it, due to the small size of the system, back action from the probe is nonnegligible;
more over, the probing result may depend on the form, the location, and the strength of the system-probe coupling, as well as the Hamiltonian and the initial state of the probe. A reliable temperature detection can be achieved only when it is free from these factors.
In the operational sense, only when such a reliable temperature detection result is achieved, can the system be regarded to possess a meaningful temperature. For small quantum systems, the operational temperature-detection method established in Ref.~\cite{WW-PRE17} is:
\bi
\item Couple a two-level probe, in an initial state $|m_0\rangle =|0\rangle$ or $|1\rangle$,
to the studied small quantum system.
\item Wait until the probe reaches a steady state $\overline{\rho}_p$ and gets a value of $\beta^{m_0}$ by fitting the steady state $\ov\rho_p$ to the Gibbs state $\frac 1Z e^{-\beta H_p}$, where $\beta^{m_0}$ is the result of $\beta$ when choosing $|m_0\rangle$ as the probe's initial state and $H_p$ is the (effective) Hamiltonian of the probe.
\item Calculate the average of $\beta^{(0)}$ and $\beta^{(1)}$, denoted by $\overline{\beta}=\frac{\beta^{(0)} +\beta^{(1)}}{2}$.
\ei

Based on the above detection method, it was found that~\cite{WW-PRE17}, for a chaotic system that is not very small in size, under an initial condition as a typical state, and within certain regime of the coupling term, the value of $\overline{\beta}$ is insensitive to details of the detection. Thus an internal temperature can be assigned to the system, with its value given by $\beta_I = \overline{\beta}$.

Now we take this method to study a small quantum system after it is coupled to a large environment and has reached a Gibbs state. We need to disconnect the system with the environment and,
 then, couple it to a two-level probe.
 The Hamiltonian of the system-probe composite is written as
\be
H_{SP}=H_{p}+\lambda H^{sp}_I +H_{S},
\ee
where $H_p=\Delta_p s_x$ and $H^{sp}_I = H^{sp}_{IS}\otimes H^{sp}_{IP}$. Here $s_x=\frac{\sigma_x}{2}$, with $\sigma_x$ being a Pauli matrix. We use $|m\ra$($m=0,1$) to denote eigenstates of the probe Hamiltonian $H_p$ with eigenvalues $e_m$, $H_p|m\ra = e_m|m\ra$. Eigenstates of the composite Hamiltonian $H_{SP}$ are denoted by $|\psi^{sp}_\mu\ra$ with energies $E^{sp}_{\mu}$, $H_{SP}|\psi^{sp} _{\mu}\rangle=E^{sp}_{\mu}|\psi^{sp} _{\mu}\rangle$, and
\begin{equation}\label{}
|\psi_{\mu}^{sp}\rangle=\sum_{k,m}C_{km}^{\mu}|\psi_{k}^{S}\rangle|m\rangle.
\end{equation}
 Let us consider $|m_0\rangle$ as the initial state of the probe.
 Based on discussions given in Sec.{~\ref{sect-ECT}}, the initial state of the system is a Gibbs state,
\be
\rho_{kk}=\frac{1}{Z}\exp(-\beta_{E}E_{k}^{S})
\ee
with
\be
Z=\sum_{k}\exp(-\beta_{E}E_{k}^{S}).
\ee

Now we turn to the evolution of the RDM of the probe. If $H_{sp}$ is chaotic, under similar consideration as in Sec.{~\ref{sect-ECT}}, the steady state of the probe should be close to its long time average $\overline{\rho}^p$, diagonalized in the eigenbasis of probe. Thus, we only need to consider the diagonal terms, which can be written as
\be
\overline{\rho}_{mm}^{(m_0)}\simeq\frac{1}{Z}\sum_{k_{0}}\exp(-\beta_{E}E_{k_{0}}^{S})\sum_{\mu}|C_{k_{0}m_{0}}^{\mu}|^{2}\sum_{k}
|C_{km}^{\mu}|^{2},
\ee
and in turn
\be\label{rho-m0mm}
\overline{\rho}_{mm}^{(m_0)}\simeq\frac{1}{Z}\sum_{k_{0}}\exp(-\beta_{E}E_{k_{0}}^{S}){\cal G}_{k_{0}m_{0}}^{m}
\ee
with ${\cal G}_{k_{0}m_{0}}^{m}=\sum_{\mu}|C_{k_{0}m_{0}}^{\mu}|^{2}G_{\mu m}$ and $G_{\mu m}=\sum_{k}|C_{km}^{\mu}|^{2}$.
Similar to the derivation of Eq.~(\ref{eq-cFkj}), if (i) $H_S$ is chaotic, and (ii) the coupling term $H^{sp}_I$ is appropriately adjusted such that $H_{sp}$ is chaotic and the width of LDOS of $H_{sp}$ is still narrow [to satisfy Eq.~(\ref{eq-LDOS-SP}) in Appendix B], we can derive ${\cal G}_{k_{0}m_{0}}^{m}$ as
\be\label{eq-Gmkm}
{\cal G}_{k_{0}m_{0}}^{m}\simeq\frac{\rho_{S}(E_{k_{0}}^{S}+e_{m_{0}}-e_{m})}{\sum_{n}\rho_{S}(E_{k_{0}}^{S}+e_{m_{0}}-e_{n})}.
\ee

In order to calculate the value of $\overline{\rho}_{mm}^{(m_0)}$, we need the concrete form of $\rho_S (E)$. Here, we consider the case that the density of states of the system  has a Gaussian form, $\rho_S (E)=\exp(-\alpha E^2)$, which is general in systems with bounded local interactions~\cite{review_Eisert16}. Moreover, one may assume that $\alpha$ can be taken as a power-law-decay function of the particle number $N$ (for the Ising model $\alpha\propto\frac{1}{N+c}$, see Ref.~\cite{GauDOS_Atas14}). From these properties,  we can get the approximate value of $\overline{\rho}_{mm}^{m_0}$ up to the third order of $\Delta_p$ (see Appendix C), i.e.,
\begin{align}
\label{eq-rhop}
\overline{\rho}_{mm}^{(m_{0})}&\simeq\frac{1}{1+\exp[2(-1)^{m}\alpha(E_{k_{0}}^{S}+e_{m_{0}})\Delta_{p}]}\nonumber\\
&+\frac{(-1)^{(1-m)}\alpha\beta\Delta_{p}^{3}}{8}.
\end{align}
Then, using $\beta^{m_{0}}=\frac{1}{\Delta_p}\ln\frac{\rho_{00}^{(m_{0})}}{\rho_{11}^{(m_{0})}}$, we have
\be
\beta^{m_{0}}\simeq\beta_{0}+2\alpha e_{m_{0}}+\frac{\alpha\beta\Delta_p ^{2}}{2}
\ee
and
\be\label{eq-betaI}
\overline{\beta}\simeq\beta_{E}+\frac{\alpha\beta_{E}}{2}\Delta_p ^{2}.
\ee
The value of $\overline{\beta}$ is independent of other factors but still depends on $\Delta_p$. Since $\alpha$ is usually a power-law-decay function of $N$, when $N$ is not small, $\overline{\beta}$ has a weak dependence on the level spacing $\Delta_p$ of the probe. Thus, $\overline{\beta}$ is insensitive to all the detection details and, hence, it can be defined as the internal temperature of the system. Namely,
\be\label{eq-betaI0}
\beta_I =\overline{\beta}.
\ee
The difference between $\beta_I$ and $\beta_E$ is therefore
\be\label{eq-dbeta}
\frac{\beta_{E}-\beta_{I}}{\beta_{E}}\simeq\frac{\alpha}{2}\Delta_p ^{2}.
\ee

At the end of this section, we discuss the difference and relation of the two concepts of temperature, i.e., internal temperature $\beta_I$ and environment-assigned temperature $\beta_E$. A system in contact with a large environment may evolve to a Gibbs state and gain its $\beta_E$ from the environment, under certain conditions stated previously. But, to possess an internal temperature usually requires more; that is, one needs to get a value of $\overline{\beta}$ insensitive to all the detail factors of the detection [see Eq.~(\ref{eq-betaI0})].

Specifically, the first requirement is that the system should be chaotic. As the second requirement, one should be able to adjust the strength of  coupling term ($\lambda$) to ensure the composite system being chaotic and the width of LDOS being narrow enough (to satisfy Eq.~(\ref{eq-LDOS-SP}) in Appendix B) at the same time. But, it is impossible to achieve this, when the system is very small. The reasons can be stated as follows. The minimum value of $\lambda$ to ensure the composite system being chaotic, denoted by $\lambda_c$ usually increase with decrease of the system's size. Thus, when the system's size is quite small, $\lambda_c$ is large.
 This usually can not guarantee narrowness of the LDOS to satisfy Eq.~(\ref{eq-LDOS-SP}) in Appendix B, which would invalid our derivations given above. Hence, a very small system may not possess an internal temperature. Another reason  is that a small $N$ usually results in a large $\alpha$, then, according to Eq.(\ref{eq-betaI}), $\overline{\beta}$ varies notably with $\Delta_p$.

%From the result of Sec.\ref{sect-ECT} and Sec.\ref{sect-INT}, it is clear that, temperature conferred by
%It can be seen from Sec.\ref{sect-ECT} that, a system can gain its temperature conferred by environment regardless of its chaotic properties and its size.
%But as we discussed in Sec.\ref{sect-INT}, more is needed for a system to possess internal temperature.
%The system should be chaotic, which determines that its size should not be very small.
%By chaotic, we means that the EFs in which the main parts of inial state is distributed are chaotic.
%Thus systems in suffiently low temperature $\beta_E$ may not have its internal temperature for the reason that EFs near the bond of spectrum usually have some regular properties, as in the case of quantum phase transition.

\section{Numerical results}\label{sect-numerical}
In this section, we put our main analytical results obtained in the above two sections into numerical  tests.

\subsection{Relaxation of the system-environment composite}
We adopt the defect Ising model as our system. It consists of $N$ spins with the nearest neighboring interactions and the periodic boundary condition. The Hamiltonian is
\be
H_{S}=\sum_{i}s_{z}^{i}s_{z}^{i+1}+g\sum_{i}s_{x}^{i}+\mu(h_{1}s_{z}^{1}+h_{4}s_{z}^{4}),
\ee
where
\be
s_{x,y,z}=\frac{1}{2}\sigma_{x,y,z}
\ee
and the parameters $h_1$, $h_4$, and $g$ are fixed ($h_1=1.11, h_4=1.61$, and $g=0.6$). The parameter $\mu$ controls  dynamical properties of the system: For $\mu=0$, the system is integrable; for $\mu\simeq[0.1,1]$, it is chaotic; and for $\mu > 1$, the system exhibits the so-called many-body localisation (MBL). In our simulations, we take $\mu=0$ and $\mu=0.3$ to represent the integrable and chaotic dynamics, respectively. The environment has a Hamiltonian
\be
H_{{\cal E}}=s_{x}^{{\cal {\cal E}}}+\lambda_{R} s_{z}^{{\cal E}}\otimes H_{IR}+H_{R},
\ee
where $H_{IR}$ is a random matrix and $H_{R}$ is a diagonal matrix, whose density of states has an exponential shape, $\rho_{R}(E) = C\exp(\beta_0 E)$. The parameter $\lambda_R$ is appropriately adjusted to ensure that the environment is chaotic. When $\lambda_R$ is not very large, $\rho_{\cal E}(E)$ has an exponential shape similar to $\rho_R (E)$ with the same exponent, i.e.,  $\rho_{{\cal E}}(E)=C'\exp(\beta_0 E)$. Here, we use $H_{\cal E}$ to simulate a chaotic environment at a temperature $\beta_0$.

\begin{figure}
\vskip-0.05cm
\includegraphics[width=1\columnwidth]{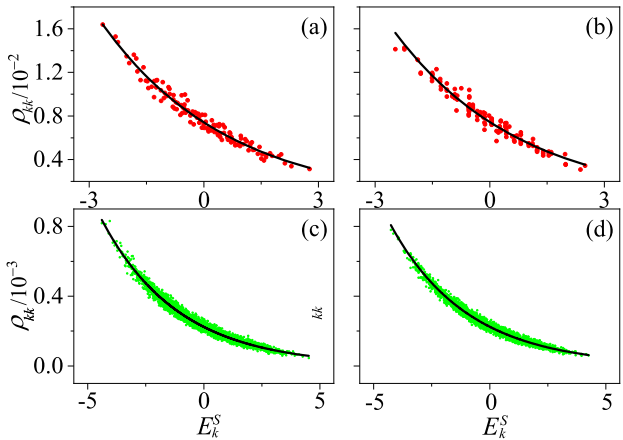}
\caption{Diagonal elements of RDM (dots) of the system, $\rho_{kk}(t)$,  versus eigenvalues $E^S _{k}$ in the defect Ising model at an instant $t$ after the system relaxed. (a) and (b) are for systems consists $N=7$ spins, with (a) chaotic case and (b) integrable case. (c) and (d) are similar to (a) and (b), respectively, but for systems consists $N=12$ spins.
The (black lines) indicate the analytical predictions by Eq.~(\ref{eq-rhokk}).
	%(a) is for chaotic system and typical state as initial state; (b) is similar to (a) but for single eigenstate as initial state;  (c) is for integrable system and typical state as initial state; (d) is similar to (c) but for single eigenstate as initial state. The black lines indicate the analytical predictions by Eq.~(\ref{eq-rhokk}).
}
\label{ECT}
\end{figure}

The coupling term $H_I$ is
\be
H_{I}=(\mu_{N}s_{z}^{N}+\mu_{N-1}s_{z}^{N-1})\otimes s_{z}^{{\cal E}}.
\ee
We set $\mu_N=0.91$ and $\mu_{N-1}=-1.11$ in our simulations in order to break the parity symmetry of the system.

To simulate the evolution of the system-environment composite, an accurate high order split-step method is used to factorize the unitary evolution operator~\cite{Num_Suzuki, Num_Casati}. In Fig.~\ref{ECT}, the results for the case that the defect Ising system has 7 and 12 spins and the environment has a dimension $d_{\cal E}=2^{12}$, are shown. It can be seen that no matter the system is chaotic or integrable, large or small, a steady RDM will be reached and it is close to the Gibbs state predicted by Eq.~(\ref{eq-rhokk}).

\begin{figure}[!]
\includegraphics[width=1\columnwidth]{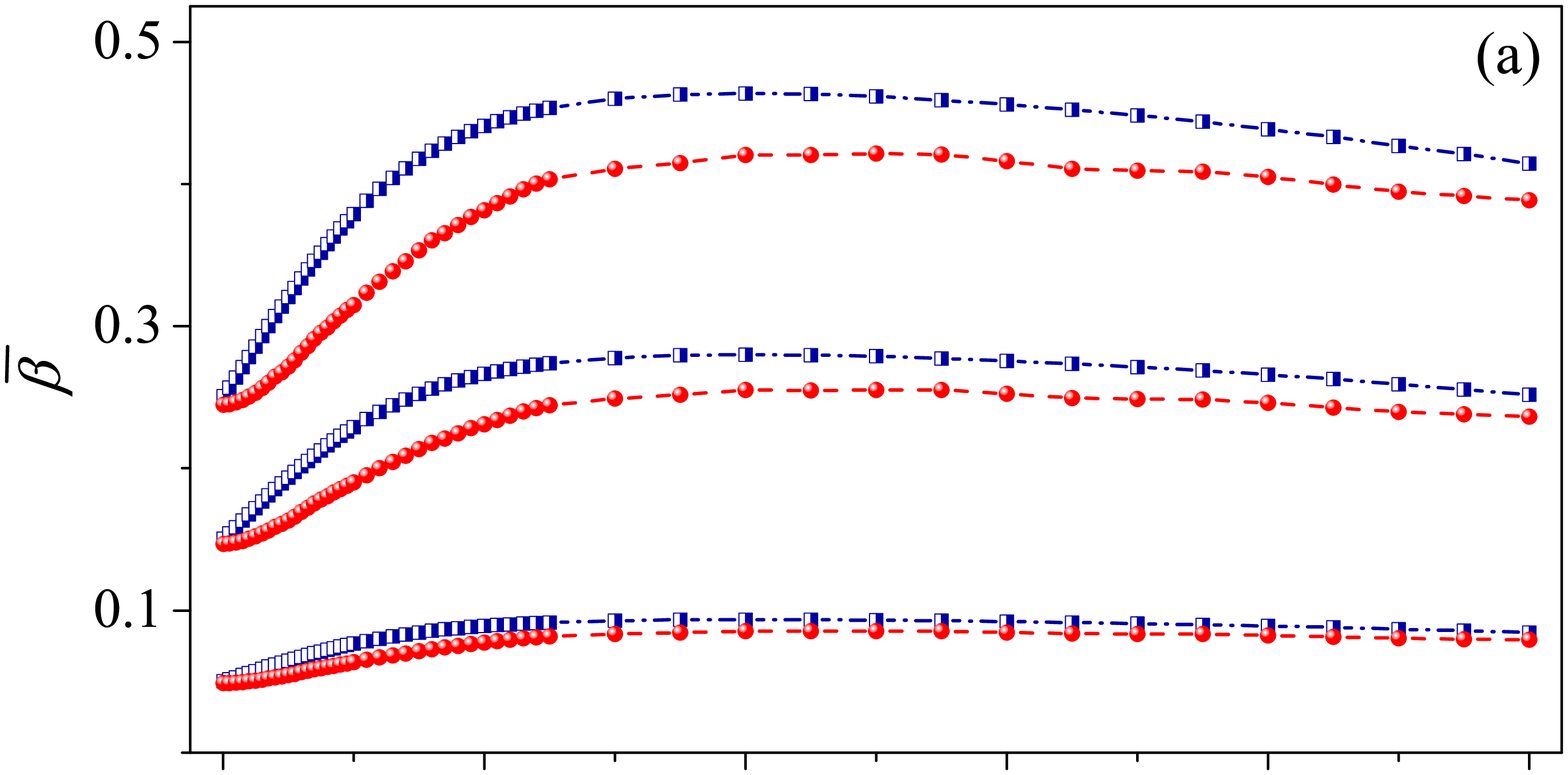}
\vskip-0.7cm
\includegraphics[width=1\columnwidth]{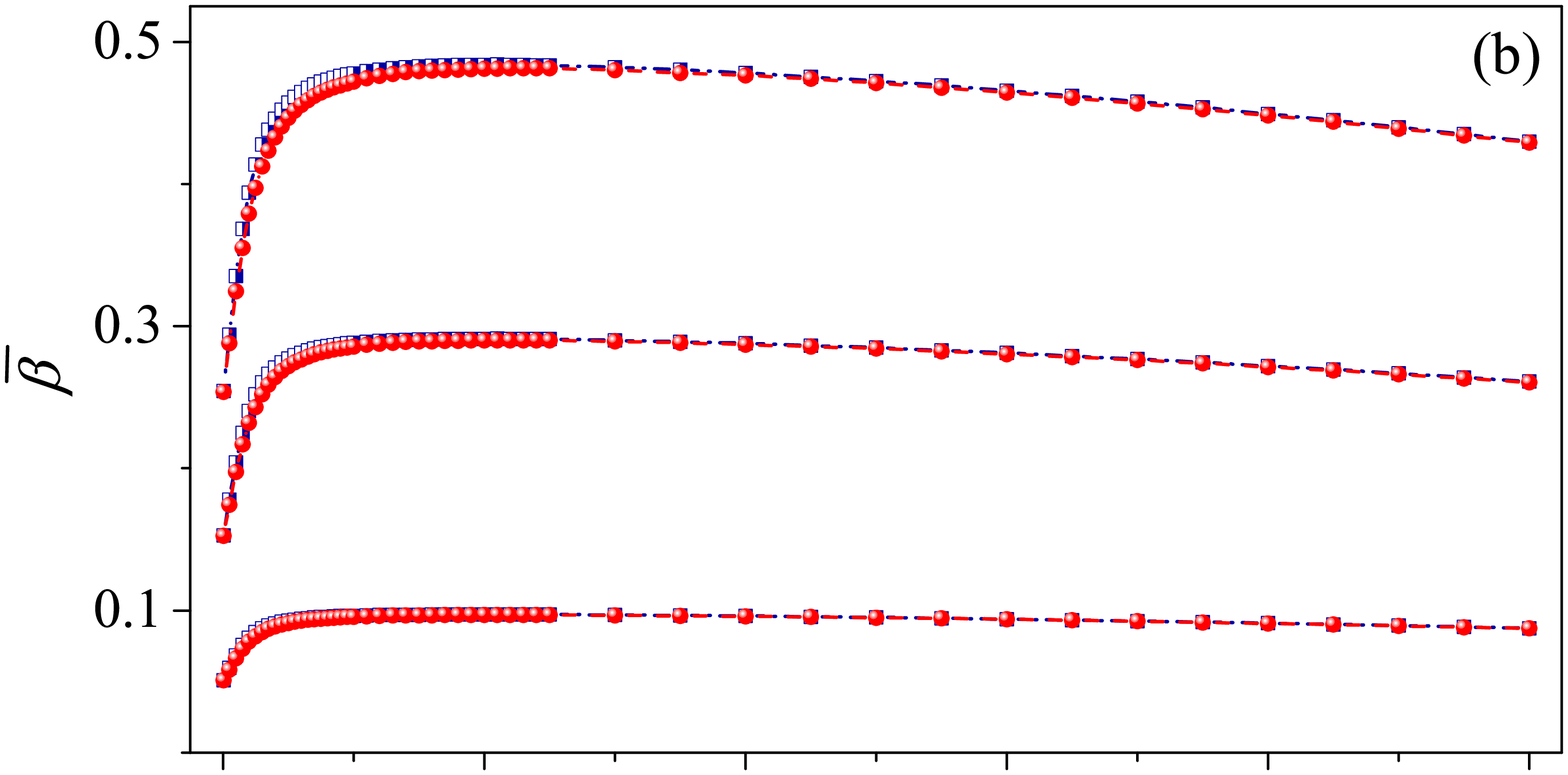}
\vskip-0.7cm
\includegraphics[width=1\columnwidth]{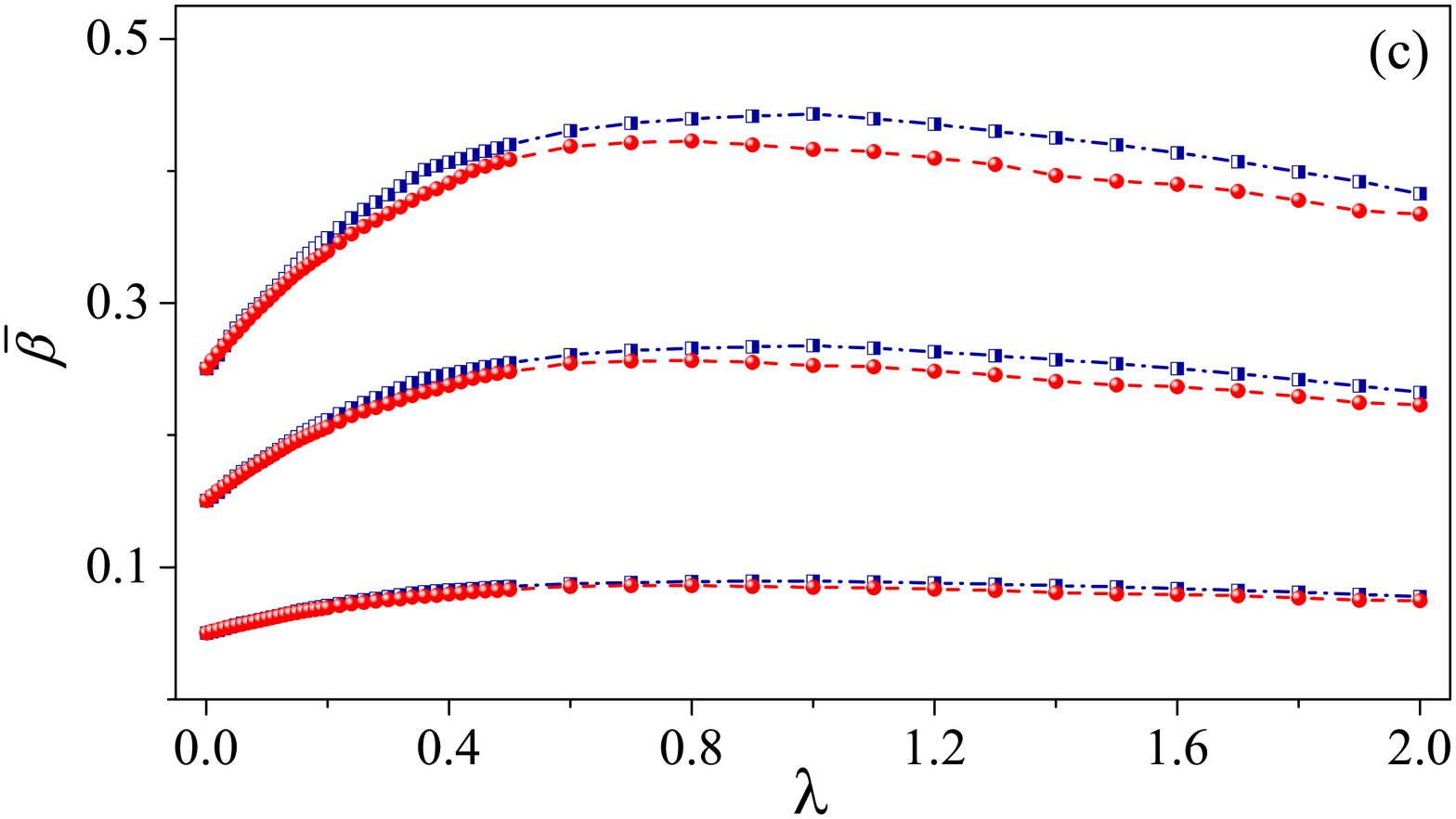}
\caption{Dependence of $\ov{\beta}$ (with $\Delta_p=0.4$) on the coupling strength $\lambda$ and the form of coupling term $H^{sp}_{IS}$ in the defect Ising model. (a) is for the integrable dynamics and the system consists of $N=13$ spins. (b) and (c) are for the chaotic dynamics with $N=13$ (b) and $N=7$ (c) spins, respectively. In each panel, and blue squares (red dots) are for $H^{sp}_{IS}=s_x$ ($H^{sp}_{IS}=s_y$); the three sets of data for a given $H^{sp}_{IS}$, from bottom to top, are for $\beta_E=0.1$, $0.3$, and $0.5$, respectively.}
\label{GibbsTchaos}
\end{figure}

\subsection{Numerical results for internal temperature of the system}
As the steady RDM of the system can be well simulated by a Gibbs state regardless of whether it is chaotic or not, in the numerical study of internal temperature of the system we use Gibbs state as its initial state instead of the true RDM.

Results of $\overline{\beta}$ for different $\lambda$ values and different form of coupling term are shown in Fig.~\ref{GibbsTchaos}. It can be seen in Fig.~\ref{GibbsTchaos}(a) that, if system is integrable, the detection result $\overline{\beta}$ depends strongly on the form of the coupling term $H^{sp}_{IS}$ and, moreover, in the whole region of $\lambda$ investigated, $\overline{\beta}$ deviates from $\beta_E$ significantly. Thus, we can conclude that  an integrable quantum system  cannot be assigned an internal  temperature.

%\begin{figure}
%\vskip-0.05cm
%\includegraphics[width=1\columnwidth]{fig3}
%\caption{Dependence of $\ov{\beta}$ on the %location (the $n$th spin), where the probe %is connected to the chaotic defect Ising %model for $\beta_E=0.1$ (red dots), %$\beta_E=0.3$ (blue squares) and %$\beta_E=0.5$ (gray triangles). The dashed %lines indicate the value of $\beta_E$ for %comparison. Here $N=13$, $\lambda=0.3$, %and $\Delta_p=0.4$.}
%\label{position}
%\end{figure}

For a chaotic system that is not small in size, as shown in Fig.~\ref{GibbsTchaos}(b) with $N=13$ spins, $\overline{\beta}$ is nearly independent of $\lambda$ in a considerably wide range of $\lambda \in(0.25,0.5)$ and the value is very close to $\beta_E$. This implies that the system has an internal temperature given by $\overline{\beta}$.
But, when the system is small, as shown in Fig.~\ref{GibbsTchaos}(c) for $N=7$, the detection result $\overline{\beta}$ depends on the form of the coupling term $H_{IS}$ and, in the whole region of $\lambda$ investigated, $\overline{\beta}$ deviates from $\beta_E$ obviously. This is in consistence with our previous analysis that a very small system does not have an internal temperature, either.

\begin{figure}[!t]
\includegraphics[width=1\columnwidth]{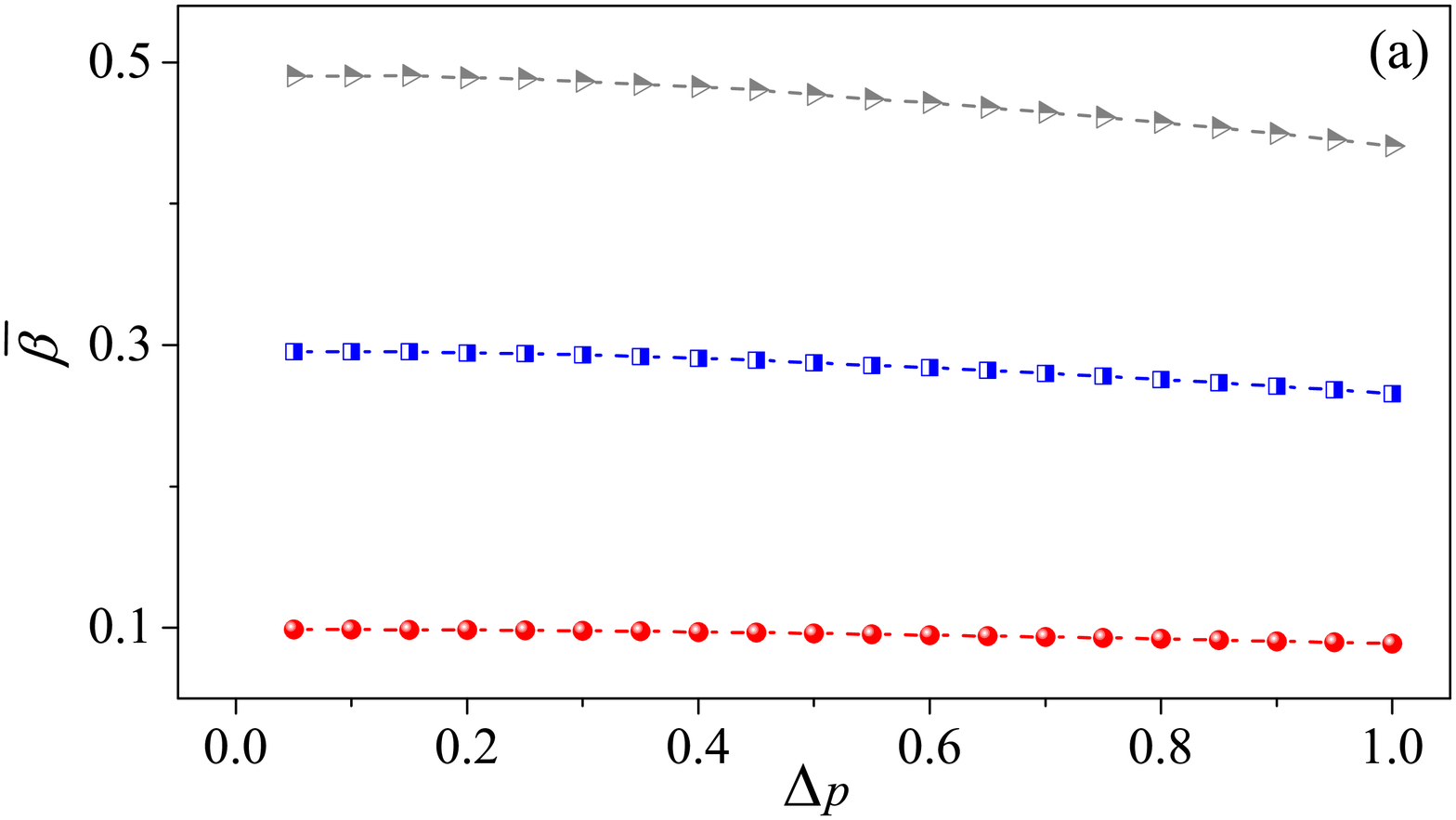}
\vskip-0.3cm
\includegraphics[width=1\columnwidth]{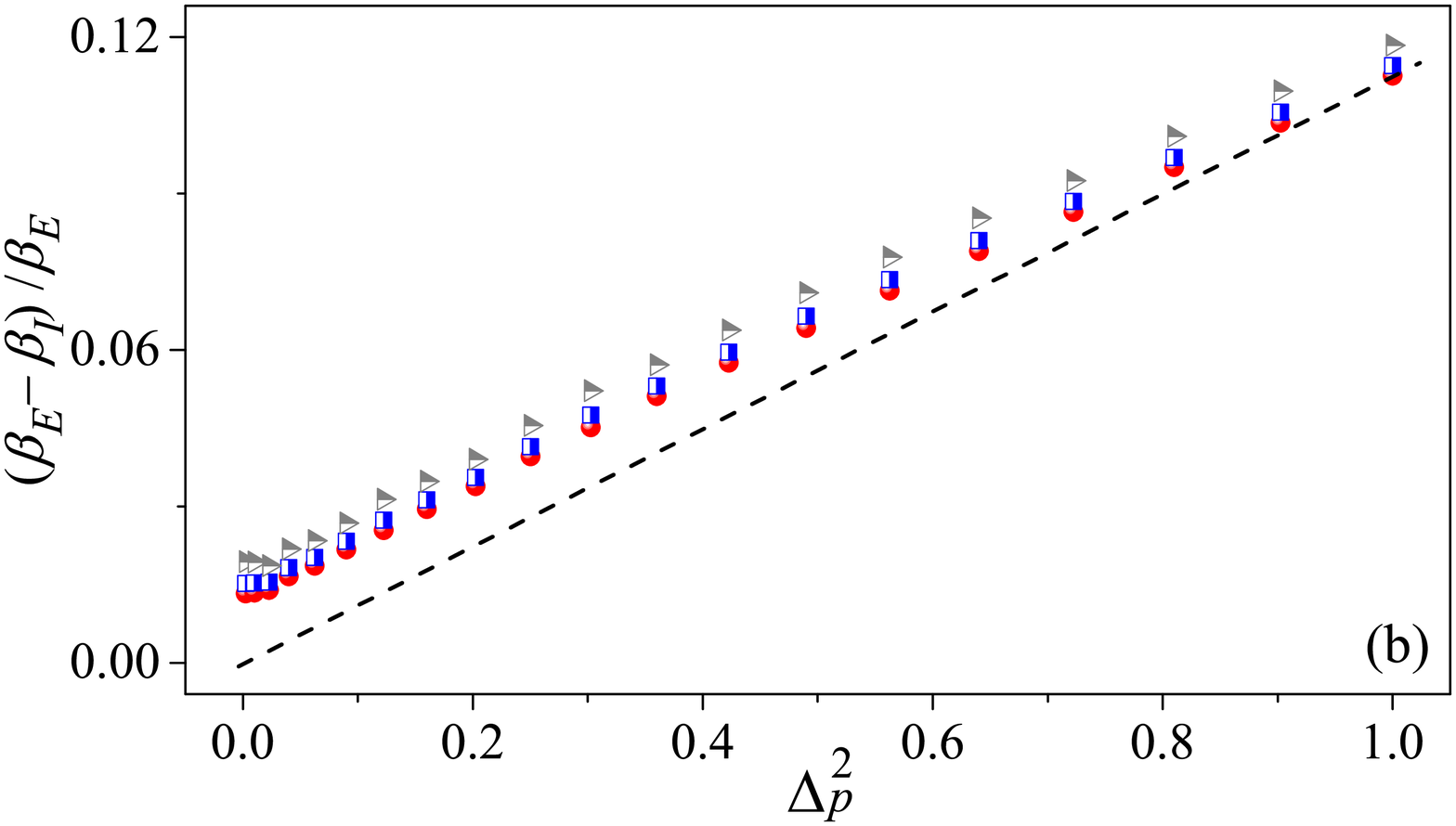}
\caption{Dependence of $\ov{\beta}$ on $\Delta_p $ (a) and on $\Delta_p ^2$
 (b), respectively, for $\beta_E=0.1$ (red dots), $\beta_E=0.3$ (blue squares) and $\beta_E=0.5$ (gray triangles). As a comparison, the black dashed line in (b) indicates our analytical result of Eq.~(\ref{eq-dbeta}).}
\label{simpledxG}
\end{figure}

Finally, in Fig.~\ref{simpledxG}(b) the validity of Eq.~(\ref{eq-dbeta}) is checked. Although it deviates obviously from the simulation results, we find that its slope agrees with that of the latter for $\Delta^2 _p\in (0,0.5)$. The main difference appears in the intercept. The reason could be that, a Gibbs state is a mix state of all eigenstates of the system, also includes those near the bounds of spectrum which have been well known to have some regular properties often and can not be regarded to be chaotic. Therefore, the value of $\overline \beta$ may deviate from our prediction of Eq.~(\ref{eq-dbeta}).
 One may introduce a small modification to Eq.~(\ref{eq-dbeta}) by
\be
\frac{\beta_{E}-\beta_{I}}{\beta_{E}}\simeq\frac{\alpha}{2}\Delta_{p}^{2}+\delta(\beta_{E}),
\ee
where $\delta(\beta_{E})$ represents a modification due to  regular eigenstates (near boundaries) in the Gibbs state. It is worth noting that the weight of the regular eigenstates usually decreases with the decrease of $\beta_E$, resulting in decrease of $\delta(\beta_E)$, which is consistent with the numerical results in Fig.~\ref{simpledxG}(b).

\section{Summary and discussions}\label{sect-concl}

In this paper, firstly, we study the condition under which small systems in contact with large, quantum chaotic environments may evolve to a Gibbs state with a parameter $\beta$ given by the environmental temperature.  It is shown that, in spite of being integrable or chaotic, large or small, such a system may evolve into the Gibbs state under appropriate interaction with the environment.
In this case, the small system gets a temperature-type property that is imposed by its environment,
which we call an environment-assigned temperature.

That a system possesses an environment-assigned temperature does not necessarily imply that
it should possess a temperature as an internal property, in the sense that it can be measured by a small probe in a reliable way.
Therefore, secondly, for a small system in a Gibbs state with an environment-assigned temperature,
we study the condition for it to possess an internal temperature in the sense mentioned above.
It is found that only chaotic systems which are not very
small may possess an internal temperature, which coincides with the environmental temperature.

Our study shows that the zeroth law of thermodynamics is still useful for small quantum systems.
But, its application may face situations much more complex than those met at the macroscopic level.
In particular, distinguishing between environment-assigned temperature and internal temperature
should be useful, in the attempt of clarifying complicated situations
that may be met when using the concept of temperature for small systems.

\acknowledgements
This work was supported by NSFC (Grants No. 11275179, No. 11535011, No. 11775210, and No. 11335006).

\begin{appendix}
\widetext
\section{Derivation of Eq.~(\ref{eq-rhokk-origin})}

In this appendix, we give the derivation of Eq.~(\ref{eq-rhokk-origin}). What is concerned here is diagonal elements $\rho_{kk} (t)$, written as
\begin{gather}
\rho_{kk}(t)=\sum_{k}\langle\phi_{j}^{{\cal E}}|\langle\psi_{k}^{S}|\exp(-iHt)|\Phi_{0}^{{\cal E}}\rangle|\psi_{0}^{S}\rangle \langle\Phi_{0}^{{\cal E}}|\langle\psi_{0}^{S}|\exp(iHt)|\phi_{j}^{{\cal E}}\rangle|\psi_{k}^{S}\rangle.
\end{gather}
Substituting $ |\Phi^{\cal E}_0\rangle = \sum_{E^{\cal E}_{j} \in \Gamma_0} D_{j} |\phi^{\cal E} _j\rangle$ and $|\psi^S _0 \rangle =\sum_k C_k |\psi ^S _k\rangle$ into the above equation, straightforward derivation gives that
\begin{gather}
\rho_{kk}(t)=\sum_{k_{0}l_{0}}C_{k_{0}}C_{l_{0}}\sum_{j_{0}i_{0}}D_{j_{0}}D_{i_{0}}\sum_{\alpha,\beta}C_{k_{0}j_{0}}^{\alpha}
C_{l_{0}i_{0}}^{\beta}\sum_{j}C_{kj}^{\alpha}C_{kj}^{\beta}\exp(-i(E_{\alpha}-E_{\beta})t),
\end{gather}
where $C^\alpha _{k_0 j_0}$ are components of the eigenfunctions (EFs) of the composite system.
 For a chaotic composite system, its energy levels are usually nondegenerate, and so do the level spacings. Therefore, for the long time average of $\rho_{kk}$, all the contributions from the terms with $\alpha \neq \beta$ can be neglected and, then, one gets that
\be
\overline{\rho}_{kk}=\sum_{k_{0}l_{0}}C_{k_{0}}C_{l_{0}}\sum_{j_{0}i_{0}}D_{j_{0}}D_{i_{0}}\sum_{\alpha}C_{k_{0}j_{0}}^{\alpha}
C_{l_{0}i_{0}}^{\alpha}\sum_{j}|C_{kj}^{\alpha}|^{2}.
\ee
It proves convenient to write $\overline{\rho}_{kk}$ as
\be
\overline{\rho}_{kk}=\overline{\rho}_{kk}^{d}+\overline{\rho}_{kk}^{nd},
\ee
where
\begin{gather}
\overline{\rho}_{kk}^{d}=\sum_{k_{0}}|C_{k_{0}}|^{2}\sum_{j_{0}}|D_{j_{0}}|^{2}\sum_{\alpha}|C_{k_{0}j_{0}}^{\alpha}|^{2}
\sum_{j}|C_{kj}^{\alpha}|^{2},\nonumber \\
\overline{\rho}_{kk}^{nd}=\sum_{k_{0}\neq l_{0} \text{ or  }j_{0}\neq i_{0}}C_{k_{0}}C_{l_{0}}D_{j_{0}}D_{i_{0}}\sum_{\alpha}C_{k_{0}j_{0}}^{\alpha}C_{l_{0}i_{0}}^{\alpha}
\sum_{j}|C_{kj}^{\alpha}|^{2}.
\end{gather}
%Here the former quantity $\overline{\rho}_{kk}^{d}$, is the summation of terms for $i_0=j_0 \text{ and } k_0=l_0$, while the latter one, $\overline{\rho}_{kk}^{nd}$, is the summation of the remaining terms.

 Below, we show that, when the dimension of the environment is sufficiently large, one has $\overline{\rho}_{kk}^{nd}\ll \overline{\rho}_{kk}^{d}$. To this end, we rewrite $\overline{\rho}_{kk}^{nd}$ as
\be \label{ov-rr}
\overline{\rho}_{kk}^{nd}=\sum_{k_{0}\neq l_{0}\text{ or  } j_{0}\neq i_{0}}C_{k_{0}}C_{l_{0}}D_{j_{0}}D_{i_{0}}S_{k_{0}l_{0}j_{0}i_{0}}^{k},
\ee
where
\be
S^k _{k_0 l_0 j_0 i_0}=\sum_{\alpha}C_{k_{0}j_{0}}^{\alpha}C_{l_{0}i_{0}}^{\alpha}\sum_{j}|C_{kj}^{\alpha}|^{2}.
\ee
Let us first study properties of $S_{k_{0}l_{0}j_{0}i_{0}}^{k}$.
For a chaotic composite system, in the basis of the uncoupled Hamiltonian $H_0$, the interaction Hamiltonian
 matrix usually has a full structure and its elements have random phases. Then, according to a result of Ref.\cite{JZ-CF}, phases of different components of the EFs have weak correlations.
 For this reason, when summing over $\alpha$, $C_{k_{0}j_{0}}^{\alpha}C_{l_{0}i_{0}}^{\alpha}$ can be regarded as random numbers.
 Note that on average $|C_{k_{0}j_{0}}^{\alpha}C_{l_{0}i_{0}}^{\alpha}|$ scales as ${1}/{D}$, where $D=d_{\cal E} d_S$ is dimension of the composite system, and  $\sum_{j}|C_{kj}^{\alpha}|^{2}\propto\frac{1}{d_{S}}$.
 Thus,
\be\label{eq-rhond}
S_{k_{0}l_{0}j_{0}i_{0}}^{k}\propto\sqrt{D}\cdot\frac{1}{D}\cdot\frac{1}{d_{S}}=\frac{1}{d_{S}\sqrt{D}}.
\ee
 Hence, $S_{k_{0}l_{0}j_{0}i_{0}}^{k}$ can be regarded as some kind of random numbers scaling as $\frac{1}{d_{S}\sqrt{D}}$.

  The coefficients $D_{j_0}$ are random numbers, scaling as $D_{j_{0}} \propto \frac{1} {\sqrt{N_{\Gamma_{0}}}} = \frac{1} {\sqrt{\rho_{{\cal E}} (E_{0}^{S})\delta E}}$.
  Moreover, $C_{k_0}$ scales as $C_{k_0}\propto \frac{1}{\sqrt{N_S}}$, where $N_S$ is the number of $k_0$ with $C_{k_0} \neq 0$. Then, from Eq.(\ref{ov-rr}), one gets the following estimate,
\be
\overline{\rho}_{kk}^{nd}\propto N_{S}N_{{\cal E}}\cdot\frac{1}{N_{S}}\cdot\frac{1}{N_{{\cal E}}}\cdot\frac{1}{d_{S}\sqrt{D}}=\frac{1}{d_{S}\sqrt{D}}=\frac{1}{d_{S}\sqrt{d_{S}d_{{\cal E}}}}.
\ee
Similarly, the scale of $\overline{\rho}_{kk}^{d}$ can be easily estimated,
\be\label{eq-rhod}
\overline{\rho}_{kk}^{d}\propto\frac{1}{d_{S}}.
\ee
From Eqs.~(\ref{eq-rhond}) and (\ref{eq-rhod}),
one finds that $\overline{\rho}_{kk}^{nd} \ll \overline{\rho}_{kk}^{d}$
when the dimension of the environment is sufficiently large.
 Hence, $\overline{\rho}_{kk}^{nd}$ can be neglected and this gives Eq.~(\ref{eq-rhokk-origin}).

\section{Derivation of Eq.~(\ref{eq-Fak}) and Eq.~(\ref{eq-cFkj})}

In this appendix, we discuss properties of the quantity $F_{\alpha k}=\sum_{j}|C_{kj}^{\alpha}|^{2}$ and ${\cal F}_{k_{0} j_{0}}^{k} = \sum_{\alpha} |C_{k_{0}j_{0}}^{\alpha}|^{2}F_{\alpha k}$.
 We derive Eqs.~(\ref{eq-Fak}) and (\ref{eq-cFkj}) under the following conditions: (i) Both the environment and the composite system are chaotic; (ii) widths of the LDOS (denoted by $w_L$) and of the  EFs (denoted by $w_E$) of the composite system are narrow. We note that the derivations to be presented below can also be used to derive Eq.~(\ref{eq-Gmkm}).

As the composite system is a quantum chaotic system, there are irregular components in the main bodies of its EFs. Making use of results given in Ref.~\cite{JZ-EF}, $C^\alpha _{kj}$ can be written in the following form
\be\label{eq-cakj}
C_{kj}^{\alpha}=R_{kj}^{\alpha}\sqrt{\langle|C_{kj}^{\alpha}|^{2}\rangle},
\ee
where $R_{kj}$ can be regarded as a Gaussian random number with unit variance and
\be
\langle|C_{kj}^{\alpha}|^{2}\rangle=\frac{1}{\rho(E_{\alpha})\epsilon}\sum_{E_{\beta}\in[E_{\alpha}-\frac{\epsilon}{2}, E_{\alpha}+\frac{\epsilon}{2}]}|C_{kj}^{\beta}|^{2},
\ee
  The quantity $\langle|C_{kj}^{\alpha}|^{2}\rangle$ gives the average shape of the EFs within
  a very narrow energy region $\epsilon$.
  It can be written as a smooth function of $E_\alpha$, i.e.,  $\langle|C_{kj}^{\alpha}|^{2}\rangle={\cal C}_{kj} (E_\alpha)$. Then, Eq.~(\ref{eq-cakj}) can be rewritten as
\be\label{eq-caEa}
C_{kj}^{\alpha}=R_{kj}^{\alpha}\sqrt{{\cal C}_{kj}(E_{\alpha})}.
\ee
Substituting Eq.~(\ref{eq-caEa}) into the expression  $F_{\alpha k}=\sum_{j}|C_{kj}^{\alpha}|^{2}$, one gets
\be
F_{\alpha k}=F_{k}(E_{\alpha})+R_{\alpha k},
\ee
where $F_k(E_\alpha)$ is a smooth part, written as
\be
F_{k}(E_{\alpha})=\sum_{j}\langle|C_{kj}^{\alpha}|^{2}\rangle=\sum_{j}{\cal C}_{kj}(E_{\alpha}),
\ee
and $R_{\alpha k}=F_{\alpha k}-F_k (E_\alpha)$ is a fluctuation part, scaling as $R_{\alpha k}\propto \frac{1}{d_S\sqrt{d_{\cal E}}}$.

To reveal properties of $F_k (E)$, we start from a Taylor expansion of $F_k (E)$, i.e.,
\be
F_{k}(E+\Delta E)\simeq \sum_{n=0}^{\infty}\frac{1}{n!}\frac{\partial^{n} F_{k}(E)}{\partial E^{n}}(\Delta E)^{n}.
\ee
 We use $\Delta^n _F$ to indicate the region of $\Delta E$, within which $F_k (E)$
 can be approximated by the above expansion cut at the $n$-th order.
  To the first order approximation, $F_k (E_\alpha)$ is calculated within the region $\Delta^1_F$
and this gives that
\begin{gather}
F_{k}(E_{\alpha})\simeq\frac{1}{N_{\alpha}}\sum_{E_{\beta}\in[E_{\alpha}-\Delta^1_F,E_{\alpha}+\Delta^1_F]}F_{k}(E_{\beta}) \simeq\frac{1}{N_{\alpha}}\sum_{E_{\beta}\in[E_{\alpha}-\Delta^1_F,E_{\alpha}+\Delta^1_F]}\sum_{j}|C_{kj}^{\alpha}|^{2},
\end{gather}
where
\be
N_{\alpha}=\sum_{E_{\beta}\in[E_{\alpha}-\Delta^1_F,E_{\alpha}+\Delta^1_F]}1
=\int_{E_{\alpha}-\Delta^1_F}^{E_{\alpha}+\Delta^1_F}\rho(E)dE.
\ee
 Here, $\rho(E)$ represents the density of states of the composite system. When the coupling is not strong, one has the following approximation,
\be
\rho(E)\simeq\sum_{l=1}^{d_{S}}\rho_{{\cal E}}(E-E_{l}^{S}),
\ee
 where $\rho_{{\cal E}}(E)$ represents the density of states of the environment.
Thus, one has
\be
N_{\alpha}\simeq\sum_{l=1}^{d_{S}}\int_{E_{\alpha}-\Delta^1_F}^{E_{\alpha}+\Delta^1_F}\rho_{{\cal E}}(E-E_{l}^{S})dE.
\ee
By introducing a step function $\Pi^{\Delta}_{E_\alpha} (E)$, defined as
\be
\Pi^{\Delta}_{E_\alpha}(E)=\begin{cases}
1 & E\in[E_{\alpha}-\Delta,E_{\alpha}+\Delta],\\
0 & otherwise,
\end{cases}
\ee
one can rewrite $F_k (E_\alpha)$ as
\be
F_{k}(E_{\alpha})=\frac{1}{N_{\alpha}}\sum_{j}\sum_{E_{\beta}}\Pi^{\Delta^1_F} _{E_\alpha}(E_\beta)|C_{kj}^{\beta}|^{2}.
\ee

Moreover, if the width of LDOS $w_L$ is narrow enough to guarantee that
\be\label{cond-LDOS}
w_L \ll \Delta^1_F,
\ee
then, for most LDOS $L_{kj}(E_\alpha)$ (with a fixed $k$ and a variant $j$), their main bodies should lie within the flat region of the function $\Pi(E)$. Taking into account  the normalization condition $\sum_\alpha |C_{km}^{\alpha}|^{2} =1$ and the fact that a narrow LDOS $L_{km}(E_\alpha)$ is approximately centered at $E_\alpha \simeq E^S _k +E^{\cal E} _j$, one finds that
\be
\sum_{E_{\beta}}\Pi_{E_{\alpha}}^{\Delta^1_F}(E_{\beta})|C_{kj}^{\beta}|^{2}\simeq\begin{cases}
1 & E_{k}^{S}+E_{j}^{{\cal E}}\in[E_{\alpha}-\Delta^1_F,E_{\alpha}+\Delta^1_F],\\
0 & \text{otherwise}.
\end{cases}
\ee
This gives that
\be\label{eq-Fk0}
F_{k}(E_{\alpha})\simeq\frac{\int_{E_{\alpha}-\Delta^1_F}^{E_{\alpha}+\Delta^1_F}\rho_{{\cal E}}(E-E_{k}^{S})dE}{\sum_{l=1}^{d_{S}}\int_{E_{\alpha}-\Delta^1_F}^{E_{\alpha}+\Delta^1_F}\rho_{{\cal E}}(E-E_{l}^{S})}.
\ee

In the case that $\Delta^1 _F$ is narrow and $\rho_{\cal E}(E)$ varies slowly within $\Delta^1 _F$, one gets the following approximation, %in Eq.~(\ref{eq-Fak}),
\be
F_{\alpha k}\simeq F_k (E_\alpha)\simeq\frac{\rho_{{\cal E}}(E_{\alpha}-E_{k}^{S})}{\sum_{l}\rho_{{\cal E}}(E_{\alpha}-E_{l}^{S})},
\ee
 which is just Eq.(\ref{eq-Fak}).
In addition, ${\cal F}_{k_{0}j_{0}}^{k}$ can be expressed using $F_k (E_\alpha)$,
\be\label{eq-cFkj0}
{\cal F}_{k_{0}j_{0}}^{k}\simeq\sum_{\alpha}|C_{k_{0}j_{0}}^{\alpha}|^{2}F_{k}(E_{\alpha}).
\ee
 If, furthermore, variation of $F_k (E_\alpha)$ within the width $w_L$ can be neglected, or
\be\label{cond-LDOS2}
w_L \le \Delta^0 _F,
\ee
then, using the normalisation condition and the fact that the LDOS  is centered at $E_\alpha \simeq E^S _k +E^{\cal E} _j$, one obtains that
\be\label{eq0}
{\cal F}_{k_{0}j_{0}}^{k}\simeq F_{k}(E_{k_{0}}^{S}+E_{j_{0}}^{{\cal E}})\simeq\frac{\rho_{{\cal E}}(E_{k_{0}}^{S}+
	E_{j_{0}}^{{\cal E}}-E_{k}^{S})}{\sum_{l}\rho_{{\cal E}}(E_{k_{0}}^{S}+E_{j_{0}}^{{\cal E}}-E_{l}^{S})},
\ee
which is just Eq.~(\ref{eq-cFkj}) in the main text.
Note that Eq.(\ref{cond-LDOS2}) usually implies Eq.(\ref{cond-LDOS}).

To summarize, the conditions that have been used to derive Eq.~(\ref{eq-Fak}) and Eq.~(\ref{eq-cFkj})
 can be rephrased as follows.
\bi
\item Both the environment and the system-environment composite are chaotic.
\item Widths of the LDOS of the composite system are sufficiently narrow, such that
\be \label{eq-LDOS-SE}
w_L \le \Delta^0 _F.
\ee
\ei
%\be\label{eq-FLDOS2}
%w_L \le \Delta^0 _F \text{ \&\& } w_L \ll \Delta^1 _{\rho_{\cal E}}.
%\ee

Noting that the above discussions do not depend on the concrete form of $\rho_{\cal E}(E)$.
 Hence, the above derivation of Eq.~(\ref{eq-cFkj}) can be extended to the derivation of Eq.~(\ref{eq-Gmkm}).
 Under similar considerations, $G_{
 \nu m}$ can be written in the following form
\be
G_{\mu m}=G_{m}(E_{\mu}^{sp})+g_{\mu m},
\ee
 where $G_{m}(E_{\mu}^{sp})$ is a smooth-varying part and $g_{\mu m}$ is a fluctuation part, scaling as $g_{\mu m}\propto\frac{1}{\sqrt{d_S}}$.
 
 We use $\Delta^n _G$ to indicate the region of $\Delta E$, within which $G_m (E+\Delta E)$
can be approximated by its Taylor expansion cut at the $n$-th order,
Then following a similar procedure, one can get the conditions to derive Eq.~(\ref{eq-Gmkm}) as follows
\bi
\item Both the system and system-probe composite are chaotic.
\item Widths of the LDOS of the system-probe composite (denoted by $w_L'$) are sufficiently  narrow, such that
\be \label{eq-LDOS-SP}
w'_{L}\le\Delta_{G}^{0}.
\ee
\ei

\section{Derivation of Eq.~(\ref{eq-rhop})}
In this section, we study properties of $\overline{\rho}^{(m_0)}_{mm}$ and derive Eq.~(\ref{eq-rhop}). It proves convenient to introduce the following quantity,
\be
{\cal G}_{m}(E)=\frac{\rho_{S}(E-e_{m})}{\sum_{n}\rho_{S}(E-e_{n})}.
\ee
As stated in the main text, we consider the case that the system's density of states has a Gaussian form, i.e., $\rho_S (E)=\exp(-\alpha E^2)$, which leads to
\be
{\cal G}_{m}(E_{k_{0}}^{S}+e_{m_{0}})=\frac{1}{1+\exp[2(-1)^{m}\alpha(E_{k_{0}}^{S}+e_{m_{0}})\Delta_{p}]}.
\ee
 Thus, Eq.~(\ref{eq-Gmkm}) can be equivalently written as
\be
{\cal G}_{k_{0}m_{0}}^{m}\simeq{\cal G}_{m}(E_{k_{0}}^{S}+e_{m_{0}}).
\ee
 Making use of this result, Eq.(\ref{rho-m0mm}), and the above expression of
 the density of states, one finds that
\begin{align}
\overline{\rho}_{mm}^{(m_{0})}&\simeq\frac{1}{Z}\sum_{k_{0}}\exp(-\beta_{E}E_{k_{0}}^{S}){\cal G}_{m}(E_{k_{0}}^{S}+e_{m_{0}}) \nonumber \\
&\simeq\frac{1}{Z}\int\rho_{S}(E)\exp(-\beta_{E}E){\cal G}(E+e_{m_{0}})dE \nonumber \\
&\simeq\frac{1}{Z}\int\exp(-\beta_{E}E-\alpha E^{2}){\cal G}(E+e_{m_{0}})dE.
\end{align}
Setting $\beta=-2\alpha E$, after some straightforward derivations, one gets that
\be
\overline{\rho}_{mm}^{(m_{0})}\simeq\frac{1}{Z}\exp(\frac{\beta_{E}^{2}}{4\alpha})\int d\beta\exp[-\frac{1}{4\alpha}(\beta-\beta_{E})^{2}]{\cal G}_{m}(-\frac{\beta}{2\alpha}+e_{m_{0}}).
\ee
Noting that 
\be
Z=\exp(\frac{\beta_{E}^{2}}{4\alpha})\int d\beta\exp[-\frac{1}{4\alpha}(\beta-\beta_{E})^{2}],
\ee
 one gets that
\be\label{eq-rhop0}
\overline{\rho}_{mm}^{(m_{0})}\simeq\frac{\int d\beta\exp(-\frac{1}{4\alpha}(\beta-\beta_{E})^{2}){\cal G}_{m}(-\frac{\beta}{2\alpha}+e_{m_{0}})}{\int d\beta\exp(-\frac{1}{4\alpha}(\beta-\beta_{E})^{2})}.
\ee
It is seen that $\overline{\rho}_{mm}^{(m_{0})}$ can be regarded as a mean value of ${\cal G}_{m} (-\frac{\beta}{2\alpha} + e_{m_{0}})$, averaged over the Gaussian distribution $\exp[-\frac{1}{4\alpha}(\beta-\beta_{E})^{2}]$. 
As stated in the main text, in many model systems with bound local interaction, the quantity $\alpha$ is a power-law decay function 
of $N$.
Hence, for sufficiently large $N$, $\overline{\rho}_{mm}^{(m_{0})}$ has the following
simple expression, 
\be
\overline{\rho}_{mm}^{(m_{0})}\simeq{\cal G}_{m}(-\frac{\beta_{E}}{2\alpha}+e_{m_{0}}).
\ee

When $N$ is not very large, we proceed from Eq.~(\ref{eq-rhop0}).
Let us focus on the case of $m=1$ and $m_0=1$. As $e_{1}=-e_{0}=\frac{\Delta_p}{2}$, one has
\be\label{eq-Gk11}
{\cal G}_{1}(-\frac{\beta}{2\alpha}+e_{1})=\frac{1}{1+\exp[\beta\Delta_p-\alpha\Delta_p ^{2}]}.
\ee
 Its Taylor expansion in $\beta$ with respect to $\beta_E$ (up to the third order) has the following form, 
\be
{\cal G}_{1}(-\frac{\beta}{2\alpha}+e_{1})\simeq{\cal G}_{1}(-\frac{\beta_{E}}{2\alpha}+e_{1})+\sum_{k=1}^{3}\frac{(\beta-\beta_{E})^{k}}{k!}\frac{\partial^{(k)}{\cal G}_{1}(-\frac{\beta}{2\alpha}+e_{1})}{\partial\beta^{k}}|_{\beta=\beta_{E}}.
\ee
Substituting the above result into Eq.~(\ref{eq-rhop0}) and noting that $\exp[-\frac{1}{4\alpha}(\beta-\beta_{E})^{2}]$ is an even function of $\beta-\beta_E$, one gets that
\be
\overline{\rho}_{11}^{(1)}\simeq{\cal G}_{1}(-\frac{\beta_{E}}{2\alpha}+e_{1})+K_{2}(\beta_{E}),
\ee
 where
\be
K_{2}(\beta_{E})=\frac{\int d\beta\exp(-\frac{1}{4\alpha}(\beta-\beta_{E})^{2})\frac{\partial^{2}{\cal G}_{1}(-\frac{\beta}{2\alpha}+e_{1})}{\partial\beta^{2}}|_{\beta=\beta_{E}}}{\int d\beta\exp(-\frac{1}{4\alpha}(\beta-\beta_{E})^{2})}.
\ee
 Making use of the expression of ${\cal G}_{1}(-\frac{\beta}{2\alpha}+e_{1})$ in Eq.~(\ref{eq-Gk11})
 and performing the integration, one finds that
\be
K_{2}(\beta_{E})=\frac{\alpha\Delta_p ^{2}\exp(\beta_{E}\Delta+\alpha\Delta_p ^{2})[\exp(\beta_{E}\Delta_p +\alpha\Delta_p ^{2})-1]}{[1+\exp(\beta_{E}\Delta_p+\alpha\Delta_p)]^{3}}.
\ee
Keeping $K_{2}(\beta_{E})$ up to the third order of $\Delta_p$, it gives
\be
K_2 (\beta_E) = \frac{\alpha\beta_E \Delta_p ^3}{8},
\ee
 and thus
\be
\overline{\rho}_{11}^{(1)}\simeq{\cal G}_{1}(-\frac{\beta_{E}}{2\alpha}+e_{1})+\frac{\alpha\beta\Delta_p ^{3}}{8}.
\ee
 Following a similar procedure, one may derive expressions for
 $\overline{\rho}_{mm}^{(m_{0})}$ of other values of $m$ and $m_0$.
  Finally, the obtained results can be written in a uniform way, i.e., 
\begin{gather}
\overline{\rho}_{mm}^{(m_{0})}\simeq{\cal G}_{m}(-\frac{\beta_{E}}{2\alpha}+e_{m_{0}})+\frac{(-1)^{1-m}\alpha\beta\Delta_p ^{3}}{8} =\frac{1}{1+\exp[2(-1)^{m}\alpha(E_{k_{0}}^{S}+e_{m_{0}})\Delta_{p}]}+\frac{(-1)^{1-m}\alpha\beta\Delta_p ^{3}}{8},
\end{gather}
 which is just Eq.~(\ref{eq-rhop}). 

\end{appendix}


\begin{thebibliography}{99}
\bibitem{Qtherm_Haar55}D. T. Haar, Rev. Mod. Phys.~\textbf{27}, 289 (1955).
\bibitem{Qtherm_Peres84} A. Peres, Phys. Rev. A~\textbf{30}, 504 (1984).
\bibitem{Qtherm_Shankar79} R. V. Jensen and R. Shankar, Phys. Rev. Lett.~\textbf{54}, 1879 (1985).
\bibitem{Qtherm_Tasaki98} H. Tasaki, Phys. Rev. Lett.~\textbf{80}, 1373 (1998).
\bibitem{Qtherm_typ_Popescu06} S. Popescu, A. J. Short, and A. Winter, {Nature Physics}~{\bf 2}, 754-758 (2006).
\bibitem{Qtherm_typ_Goldstein06} S. Goldstein, J. L. Lebowitz, R. Tumulka, and N. Zanghi, {Phys. Rev. Lett.}  {\bf 96}, 050403
        (2006).
\bibitem{Qtherm_Neuann} J. v. Neumann, Zeitschrift für Physik~\textbf{57}, 30 (1929), English translation (by R. Tumulka), The
         European Physical Journal H 35, \textbf{201} (2010).

\bibitem{Exp_Kinoshita06} T. Kinoshita, T. Wenger, and D. S. Weiss, Nature~\textbf{440}, 900 (2006).
\bibitem{Exp_Gring12} M. Gring, M. Kuhnert, T. Langen, T. Kitagawa, B. Rauer, M. Schreitl, I. Mazets, D. A. Smith, E. Demler,
and J. Schmiedmayer, Science~\textbf{337}, 1318 (2012).
\bibitem{Exp_Gross12} M. Cheneau, P. Barmettler, D. Poletti, M. Endres,
P. Schauß, T. Fukuhara, C. Gross, I. Bloch, C. Kollath,
and S. Kuhr, Nature~\textbf{481}, 484 (2012).
\bibitem{Exp_Bloch12} U. Schneider, L. Hackerm\"{u}ller, J. P. Ronzheimer, S. Will,
S. Braun, T. Best, I. Bloch, E. Demler, S. Mandt, D. Rasch, et al., Nature Physics~\textbf{8}, 213 (2012).

\bibitem{Exp_Eisert12} S. Trotzky, Y.-A. Chen, A. Flesch, I. P. McCulloch,
U. Schollw\"{o}ck, J. Eisert, and I. Bloch, Nature Physics~\textbf{8}, 325 (2012).
\bibitem{Exp_Langen13} T. Langen, R. Geiger, M. Kuhnert, B. Rauer, and
J. Schmiedmayer, Nature Physics~\textbf{9}, 640 (2013).
\bibitem{Exp_Greiner16} A. M. Kaufman, M. E. Tai, A. Lukin, M. Rispoli,
R. Schittko, P. M. Preiss, and M. Greiner, Science~\textbf{353}, 794 (2016).
\bibitem{Exp_Richerme14} P. Richerme, Z.-X. Gong, A. Lee, C. Senko, J. Smith,
M. Foss-Feig, S. Michalakis, A. V. Gorshkov, and C. Monroe, Nature~\textbf{511}, 198 (2014).
%\bibitem{Exp_Roos14} P. Jurcevic, B. P. Lanyon, P. Hauke, C. Hempel, P. Zoller,
%R. Blatt, and C. F. Roos, Nature \textbf{511}, 202 (2014).
\bibitem{Exp_Clos16} G. Clos, D. Porras, U. Warring, and T. Schaetz, Phys.
Rev. Lett.~\textbf{117}, 170401 (2016).
%\bibitem{Exp_Huse16} J. Smith, A. Lee, P. Richerme, B. Neyenhuis, P. W. Hess,
%P. Hauke, M. Heyl, D. A. Huse, and C. Monroe, Nature Physics \textbf{12}, 907 (2016).

\bibitem{review_Polkovnikov16}A. Polkovnikov and D. Sels, Science~\textbf{353}, 752 (2016).
\bibitem{review_Polkovnikov11}A. Polkovnikov, K. Sengupta, A. Silva, and M. Vengalattore, Rev. Mod. Phys.~\textbf{83}, 863 (2011).
\bibitem{review_Eisert15} J. Eisert, M. Friesdorf, and C. Gogolin, Nature Physics~\textbf{11}, 124 (2015).

\bibitem{review_Rigol16} L. D’Alessio, Y. Kafri, A. Polkovnikov, and M. Rigol, Advances in Physics~\textbf{65}, 239 (2016).
\bibitem{review_Izrailev16}F. Borgonovi, F. M. Izrailev, L. F. Santos, V. G. Zelevinsky, Phys. Rep.~\textbf{626}, 1-58 (2016).

\bibitem{ETH_Deutsch} J. M. Deutsch, Phys. Rev. A~\textbf{43}, 2046 (1991).
\bibitem{ETH_Srednicki} M. Srednicki, Phys. Rev. E~\textbf{50}, 888 (1994).

\bibitem{ETH_Rigol08} M. Rigol, V. Dunjko, and M. Olshanii, Nature~\textbf{452}, 854 (2008).

\bibitem{Qtherm_Iso_Srednicki99} M. Srednicki, J. Phys. A~\textbf{32} 1163 (1999).
\bibitem{Qtherm_Iso_Izrailev12} L. F. Santos, F. Borgonovi, and F. M. Izrailev,	Phys. Rev. Lett.~\textbf{108} 094102 (2012).
\bibitem{Qtherm_Iso_Rigol12} M. Rigol and M. Srednicki, Phys. Rev. Lett.~\textbf{108}, 110601 (2012).
\bibitem{Qtherm_Iso_Rigol14} M. Rigol, Phys. Rev. Lett.~\textbf{112}, 170601 (2014).
\bibitem{Qtherm_Iso_Sorg14}S. Sorg, L. Vidmar, L. Pollet, and F. Heidrich-Meisner, Phys. Rev. A~\textbf{90}, 033606 (2014).
\bibitem{Qtherm_Iso_Santos14} E. J. Torres-Herrera and L. F. Santos, Phys. Rev. E~\textbf{89}, 062110 (2014).




\bibitem{Qtherm_SE_Yuan09}S. Yuan, Mikhail I. Katsnelson, and H. De Raedt, J. Phys. Soc. Jpn.~\textbf{78}, 094003 (2009).
\bibitem{Qtherm_SE_Linden09} N. Linden, S. Popescu, A. J. Short, and A. Winter, {Phys. Rev. E}~\textbf{79}, 061103 (2009).
\bibitem{Qtherm_SE_Genway12} S. Genway, A. F. Ho, and D. K. K. Lee, Phys. Rev. A~\textbf{86}, 023609 (2012).
\bibitem{Qtherm_SE_Eisert12} A. Riera, C. Gogolin, and J. Eisert, Phys. Rev. Lett.~\textbf{108}, 080402 (2012).
\bibitem{Qtherm_SE_WWG12} W.-g.~Wang, Phys. Rev. E~\text{86}, 011115 (2012).
\bibitem{Sed13} N. Sedlmayr, J. Ren, F. Gebhard, and J. Sirker, Phys. Rev. Lett. \textbf{110}, 100406 (2013).
\bibitem{Qtherm_SE_Freeman14} D. Freeman, C. M. Herdman, D. J. Gorman, and K. B. Whaley, Phys. Rev. B~\textbf{90}, 134302 (2014).
\bibitem{Qtherm_SE_Cramer15}G. De Palma, A. Serafini, V. Giovannetti, and M. Cramer. Phys. Rev. Lett.~\textbf{115}, 220401 (2015)
\bibitem{Qtherm_SE_Fiako15} O. Fialko, Phys. Rev. E~\textbf{92}, 022104, (2015).




\bibitem{Temp_Hanggi14} S. Hilbert, Peter H\"{a}nggi, and J. Dunkel, Phys. Rev. E~\textbf{90}, 062116 (2014).
\bibitem{Temp_local_Eisert14} M. Kliesch, \textit{et. al}, Phys. Rev. X~\textbf{4}, 031019 (2014).
\bibitem{Temp_NET_Braun13} S. Braun, \textit{et al.}, Science \textbf{339}(6115), 52-55, (2013).

\bibitem{WW-PRE17} J. Wang and W.-g. Wang, Phys. Rev. E \textbf{96},032207 (2017).
\bibitem{pre14-ps} L. He and W.-g. Wang, Phys.Rev.E~\textbf{89}, 022125 (2014).
\bibitem{YWW19} H. Yan, J. Wang, and W.-g. Wang, arXiv:1902.10944.
\bibitem{Zurek-ps}  J. P. Paz and W. H. Zurek, Phys. Rev. Lett.~\textbf{82}, 5181 (1999).
\bibitem{pra08-ps} W.-g. Wang, J. Gong, G. Casati, and B. Li, Phys.~Rev.~A~\textbf{77},
 012108 (2008).
\bibitem{review_Eisert16} C. Gogolin and J. Eisert, Rep. Prog. Phys.~\textbf{79}, 056001 (2016).
\bibitem{GauDOS_Atas14} Y. Y. Atas and E. Bogomolny, J. Phys. A \textbf{47}, 335201 (2014).
\bibitem{Num_Suzuki} M. Suzuki,Phys. Lett. A~\textbf{165}, 387 (1992).
\bibitem{Num_Casati} C. Mejia-Monasterio, T. Prosen, and G. Casati, Europhys. Lett.~\textbf{72}, 520 (2005).


%\bibitem{chaosEF_WW} Jiaozi wang and Wen-ge Wang, in prepare.






\end{thebibliography}

\begin{thebibliography}{99}
\bibitem{JZ-CF}  Jiaozi Wang and Wen-ge Wang, Phys. Rev E .\textbf{96}, 052221 (2017).
	
\bibitem{JZ-EF}  Jiaozi Wang and Wen-ge Wang,  Phys. Rev. E \textbf{97}, 062219 (2018).
	
\end{thebibliography}
\end{document}